\documentclass[useAMS,usenatbib]{mn2e}
\usepackage[dvips]{graphicx}
\usepackage{graphicx}

\title[Segregation of Two Component Advective Flow]
{Segregation of a Keplerian disc and sub-Keplerian halo from a Transonic flow around a Black Hole by Viscosity and 
Cooling processes}
\author[Kinsuk Giri, Sudip K. Garain and Sandip K. Chakrabarti]
{Kinsuk Giri\thanks{kinsuk@mx.nthu.edu.tw}$^{1}$, Sudip K. Garain\thanks{sgarain@nd.edu}$^{2}$
 and Sandip K. Chakrabarti\thanks{chakraba@bose.res.in}$^{3,4}$\\
$^{1}$Institute of Astronomy, National Tsing Hua University, 101 Section 2 Kuang Fu Road, Hsinchu 30013, Taiwan, ROC \\
$^{2}$Department of Physics, University of Notre Dame, Notre Dame, IN 46556, USA \\
$^{3}$S. N. Bose National Centre for Basic Sciences, Salt Lake, Kolkata 700098, India\\
$^{4}$Indian Centre for Space Physics, Chalantika 43, Garia Station Rd., Kolkata, 700084, India}
\begin{document}
\date{}

\maketitle

\label{firstpage}
\begin{abstract}
A black hole accretion is necessarily transonic. In presence of sufficiently high viscosity and 
cooling effects, a low-angular momentum transonic flow can become a standard Keplerian disc except close to the
where hole where it must pass through the inner sonic point. However, if the viscosity is not 
high everywhere and cooling is not efficient everywhere, the flow cannot completely become a Keplerian disc. 
In this paper, we show results of rigorous numerical simulations of a transonic flow having vertically 
varying viscosity parameter (being highest on the equatorial plane) and optical depth dependent cooling 
processes to show that the flow indeed segregates into two distinct components as it approaches a black hole.
The component on the equatorial plane has properties of a standard Keplerian disc, though the flow
is not truncated at the innermost stable circular orbit. This component extends till the horizon as a sub-Keplerian 
flow. This standard disc is found to be surrounded by a hot, low angular momentum component forming 
a centrifugal barrier dominated oscillating shock wave, consistent with the Chakrabarti-Titarchuk two component advective flow 
configuration.
\end{abstract}


\section{Introduction}

It is well known that a standard Keplerian disc (Shakura \& Sunyaev, 1973, hereafter SS73; 
Novikov \& Thorne, 1973) is not capable of explaining the entire X-ray and Gamma-ray spectrum characterizing 
an accreting black hole candidate. This is because SS73 disc emits a multicolour black body
spectrum whereas a general observed spectrum contains an additional power-law component (e.g., Sunyaev \& Truemper,  1979).
Though there are several ad hoc models to explain power-law component 
(generally considered to arise out of Comptonization of soft photons by a hot electron cloud), 
Chakrabarti \& Titarchuk (1995, hereafter CT95) explained the spectra using a truly global {\it solution}
in which a sub-Keplerian (low-angular momentum) flow which produces a centrifugal barrier 
close to a black hole, and surrounds an SS73 disc (corrected at the inner edge), plays a major role. 
This so-called Two Component Advective Flow (TCAF) solution was taken straight out of 
theoretical study of the behaviour of topology of viscous flows around black holes (Chakrabarti, 1990, 1996). 
In a major breakthrough in understanding properties of a viscous transonic flow, it was pointed out (Chakrabarti 1990, 1996)
that flows with super-critical viscosity parameter ($\alpha > \alpha_{crit}$) everywhere can only form a Keplerian disc. 
In CT95, precisely this property of the flow was used. One component with
a $\alpha>\alpha_{crit}$ is a cooler, Keplerian disc on the equatorial plane, 
accreting in a viscous time scale. The second component is a low angular momentum flow with 
sub-critical ($\alpha<\alpha_{crit}$) viscosity parameter which forms a centrifugal pressure supported standing or oscillating shock 
and is accreting in almost a free-fall time scale. Later, many workers including Smith et al. 
(2001a), Smith et al. (2001b), Miller et al. (2001), Soria et al. (2009, 2011), Dutta \& Chakrabarti (2010), Cambier \& Smith (2013)
found evidences of two components in many of the black hole candidates 
which motivated us to investigate into this configuration further.

Since TCAF claims to be the only theoretical solution which can describe 
the whole system, including hydrodynamic (disc plus outflows),
spectral and temporal properties, its origin and stability are to be established by rigorous means. In a series of papers
(Giri et al. 2010, hereafter Paper I; Giri \& Chakrabarti, 2012, hereafter Paper II; Giri \& Chakrabarti, 2013, hereafter III), 
it was shown how a viscous flow may produce a Keplerian disc on the equatorial plane if viscosity parameter is 
super-critical. Away from the equatorial plane, where viscosity parameter could be smaller, angular momentum
distribution remains sub-Keplerian. A toy model was tried out in Paper III, 
where the entire flow was treated with a power-law cooling (not to be confused with the power-law spectrum produced 
due to Comptonization as mentioned earlier.)
so that denser Keplerian disc could cool down faster while lighter sub-Keplerian halo remains hot and produces 
the CENtrifugal pressure dominated BOundary Layer or CENBOL, exactly as envisaged in CT95. 
However, a standard SS73 disc not only has a Keplerian distribution, it also emits a black 
body spectrum at a local temperature. So it needs to cool faster to emit such a radiation.
In the present paper, we concentrate our attention 
to establish TCAF solution with hydrodynamic and radiative properties exactly as envisaged originally.  
We use the same time dependent hydrodynamic code as in Paper III, but instead of using a 
power-law cooling process (as a proxy to Comptonization which is stronger than the
normal bremmstrahlung cooling) 
throughout, we use cooling processes which depend on optical depths at a given point. This process
is expected to produce an SS73 type disc where optical depth is high and leave the 
lower optical depth region radiatively inefficient as in a low-angular momentum transonic flow.

\section{Basic Equations anid Cooling Laws}

Basic equations governing a two-dimensional axisymmetric flow around a Schwarzschild 
black hole have been presented extensively in earlier papers (Papers I-III) and we 
describe them here very briefly. Cylindrical coordinate system  $(r, \phi, z)$ is adopted with 
z-axis being rotation axis of the disc. We use mass of the black hole $M_{BH}$, the velocity of light $c$ and 
Schwarzschild radius $R_g=2GM_{BH}/c^2$ as units of mass, velocity and distance respectively.  
Viscosity prescription has been used in a similar way as in Paper III. 
However, unlike Paper III, we incorporate cooling law differently.
In each time step of our simulation, we use two types of coolings depending 
on optical depth of matter at a given computational grid.

At the beginning of our simulation, we start with a power-law cooling (as before) having
a cooling rate of $$\Lambda_{br}=1.4\times 10^{-27}N^2(r,z)T^{\beta}(r,z) {\rm~ erg~cm^{-3}~s^{-1}},   \eqno(1)$$
where, $N(r,z)$ is the number density and $T(r,z)$ is temperature of the flow at a location $(r,z)$. 
This power-law cooling was used as a proxy for Comptonization which is stronger than bremsstrahlung.
For $\beta=0.5$, this becomes the well known bremsstrahlung cooling. 
Gas temperature $T$ at any ($r,z)$ is obtained from density ($\rho$) and
pressure ($p$) obtained from our simulation using the ideal gas law: 
$$
T = {p \over \rho}{{\mu {m_p}} \over {k_b}},  \eqno(2)
$$
where, $k_b$ is Boltzmann constant, $\mu = 0.5$, for purely hydrogen gas.    
Initially (transient period of the simulation) to increase the cooling efficiency,
we took cooling index to be $\beta = 0.6$ (as in Molteni, Sponholz \& Chakrabarti, 1996).
This was done to initiate the cooling process. At each time step, we compute optical depth, 
$$\tau=\int N(r,z)\sigma_T dz, \eqno(3)$$ 
where $\sigma_T$ is the Thomson scattering cross-section, along
vertical directions starting from upper grid boundary to the equatorial plane. As soon as
a Keplerian disc starts forming near the equatorial plane, we find a sudden
increase in $\tau$ and we define the height
where $\tau$ changes abruptly to be the surface of the disc. This surface is dynamically 
detected by the code. In case of an optically thick, geometrically thin
Keplerian disc, energy is assumed to be radiated from the surface in the form of a 
black body radiation, with a cooling rate of $$\Lambda_{bb} = \sigma T^4(r,z) {\rm~erg~cm^{-2}~s^{-1}},  \eqno(4)$$
where, $\sigma$ is Stefan-Boltzmann constant given by $\sigma=5.67\times 10^{-5}~
{\rm erg~cm^{-2}~deg^{-4}~s^{-1}}$. 
In regions below the surface of the disc, a separate cooling is no longer necessary as the energy produced
in this region in expense of gravitational energy of the flow, is either diffused as above
or advected into another grid. It is to be noted that advection is important in 
this component only close to the inner edge, where optical depth is never high 
enough to emit as a black body. Advection is also important in the 
sub-Keplerian region. In these two cases, cooling remains as in $\Lambda_{br}$ (Eqn. 1).

\section{Setup of the Problem}
Setup of our simulation is the same as presented in Paper III. We repeat only some key points. 
As before, we assume that gravitational field of 
the black hole can be described by Paczy\'{n}ski \& Wiita (1980) pseudo-Newtonian potential,
$$
\phi(r,z) = -{GM_{BH}\over(R-R_g)},     \eqno(5)
$$
where, $R=\sqrt{r^2+z^2}$.
We assume a polytropic equation of state for 
accreting (or, outflowing) matter, $P=K \rho^{\gamma}$, where,
$P$ and $\rho$ are the isotropic pressure and matter density
respectively, $\gamma$ is a constant throughout the flow and $K$ is related
to specific entropy of the flow $s$ (which need not be constant). 
In all our simulations, where the flow expected to be relativistic, we take $\gamma = 4/3$.

The computational box occupies one quadrant of r-z plane with $0 \leq r \leq 200$ 
and $0 \leq z \leq 200$. 
We use reflection boundary condition on equatorial plane and z-axis to obtain a solution in other quadrants. 
Incoming gas enters the box through outer boundary 
located at $r_b = 200$. We have chosen density of incoming gas ${\rho}_{in} = 2.5$ for convenience. 
We supply radial velocity $v_r$ and sound speed $a$ of flow at outer boundary grids parallel to 
z-axis (at $r=r_b$). We take boundary values of density from standard vertical equilibrium solution 
of a transonic flow (Chakrabarti, 1989). We inject matter at outer boundary placed along z-axis. 
In order to the mimic horizon of the black hole, we place an absorbing inner boundary at $R = 2.5 R_g$, 
inside which all the matter is completely absorbed. Initially, the grid was filled with a background 
matter (BGM) of density ${\rho}_{bg} = 10^{-6}$ having a sound speed (or, temperature) 
same as that of the injected matter. Hence, injected matter has $10^6$ times larger pressure 
than that of the BGM. The BGM is placed to avoid non-physical singularities caused by 
`division by zero'. This matter is quickly washed out and replaced by injected matter within a dynamical time scale. 
Calculations were performed with $512 \times 512$ cells, so that each grid has a size of $0.39$ in units of 
Schwarzschild radius. Free fall timescale from the outer boundary is about $0.43$s as computed from the sum 
of $dr/<v_r>$ over the entire radial grid, $<v_r>$ being averaged over $20$ vertical grids.

All simulations have been carried out assuming a stellar mass black hole $(M_{BH} = 10{M_\odot})$. 
We carry out simulations for several hundreds of dynamical time-scales. In reality, our simulation time 
corresponds to a few seconds in physical units. Conversion factor of our time unit to physical unit
is $2GM_{BH}/c^3$, and thus physical time for which the program was run would scale with the mass of the black hole.

\section{Simulation Results}

We numerically solved system of equations discussed in Paper III and above,
using a grid based finite difference method which is called
Total Variation Diminishing (TVD) technique to deal with laboratory fluid dynamics (Harten, 1983) and 
was further developed by Ryu et al. (1995 \& 1997) to study non-viscous astrophysical flows without cooling.
We assume accretion flow to be in vertical equilibrium  
at the outer boundary so that we can inject matter according to theoretical transonic solution (Chakrabarti, 1989). 
The injection rate of momentum density is kept uniform throughout injected height
at the outer boundary. Simulation is run till $t = 95$ s, which is
more than two hundred times the dynamical time.
Thus, solutions presented by us are obtained long after the transient phase is over in about a second.
We have run several cases.  In Table 1, we show parameters used for the simulations at a glance.
The case IDs are given in the first column. Columns 2 and 3 show specific energy ($\cal E$, energy per unit mass) 
and specific angular momentum ($\lambda$) of injected flow at outer boundary. The energy
and the angular momentum are measured in the unit of $c^2$ and $r_gc$, respectively.
In column 4, we present the injection rate of matter ($\dot{M}$) at the outer boundary. 
It is measured in the units of mass Eddington rate ($1.44 \times 10^{17}M_{BH}$ gm/s).
Finally, in column 5, we put values of viscous parameter, a dimensionless number $\alpha$.
Results of our simulation are discussed below. 
\begin{table*}
\addtolength{\tabcolsep}{-1.0pt}
\begin{center}
\begin{tabular}{|c|c|c|c|c|c|c|c|c|c|c|c|c|}
\hline
\hline Case ID & ${\cal E}$ & $\lambda$ & $\dot{M}$ & $\alpha$ \\
\hline C1 & $0.001$ & $1.70$ & $2.0$ & $0.012$ \\
\hline C2 & $0.001$ & $1.70$ & $1.5$ & $0.012$ \\
\hline C3 & $0.001$ & $1.70$ & $1.0$ & $0.012$ \\
\hline C4 & $0.001$ & $1.70$ & $0.5$ & $0.012$ \\
\hline C5 & $0.001$ & $1.95$ & $2.0$ & $0.012$ \\
\end{tabular}
\end{center}
\caption{Parameters used for the simulations at a glance}
\end{table*}
In Fig. 1, we compare simulated specific angular momentum distributions throughout $r-z$ plane  and corresponding Keplerian 
distribution for the case {\bf C1}. This is a typical input for a transonic flow which we wish to convert to a Keplerian disc. 
In Fig. 1a, the result obtained from our simulation is shown.   
\begin{figure}
\begin{center}
\includegraphics[height=12truecm,width=10truecm,angle=0]{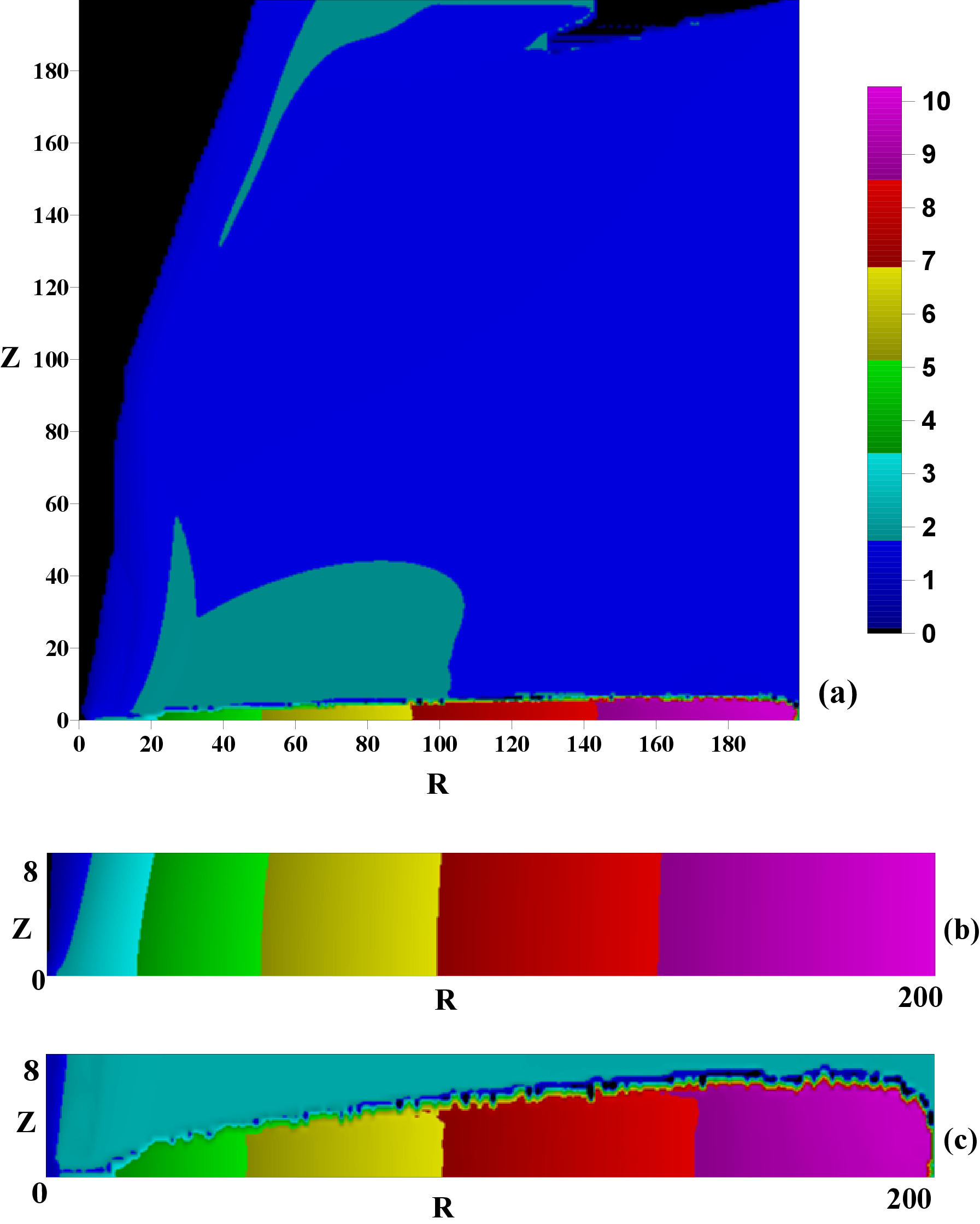}
\caption{(a) Specific angular momentum distribution at $r-z$ plane in our simulation (Top panel) for {\bf C1}.
To compare with analytical results, region close to the equatorial plane is zoomed in to show
(b) distribution in our simulated result and (c) Keplerian distribution.}
\end{center}
\end{figure}
In order to focus on our simulated Keplerian disc, in Fig. 1b, we zoom regions close to the equatorial plane while the same
is shown in Fig. 1c for theoretical Keplerian distribution. Distributions in Fig. 1 (a-c) are plotted in linear scale. 
Different colours (online version) correspond to different specific angular momentum as marked in scale on right. 
It is to be noted that the outer front of Keplerian disc moves very slowly as compared to the inflow velocity 
of sub-Keplerian matter. In the frame of sub-Keplerian matter, the Keplerian disc behaves as an obstacle. 
This causes formation of an wake at the tip of the outer edge of simulated Keplerian flow. In a real situation, 
Keplerian disc will be extended till the outer boundary and Keplerian
matter would be injected in those grids instead of a transonic flow. Our explicit procedure given here only demonstrates
how a Keplerian disc forms in the first place, not by removal, but by redistribution of injected angular momentum.
\begin{figure}
\begin{center}
\includegraphics[height=13truecm,width=9truecm,angle=0]{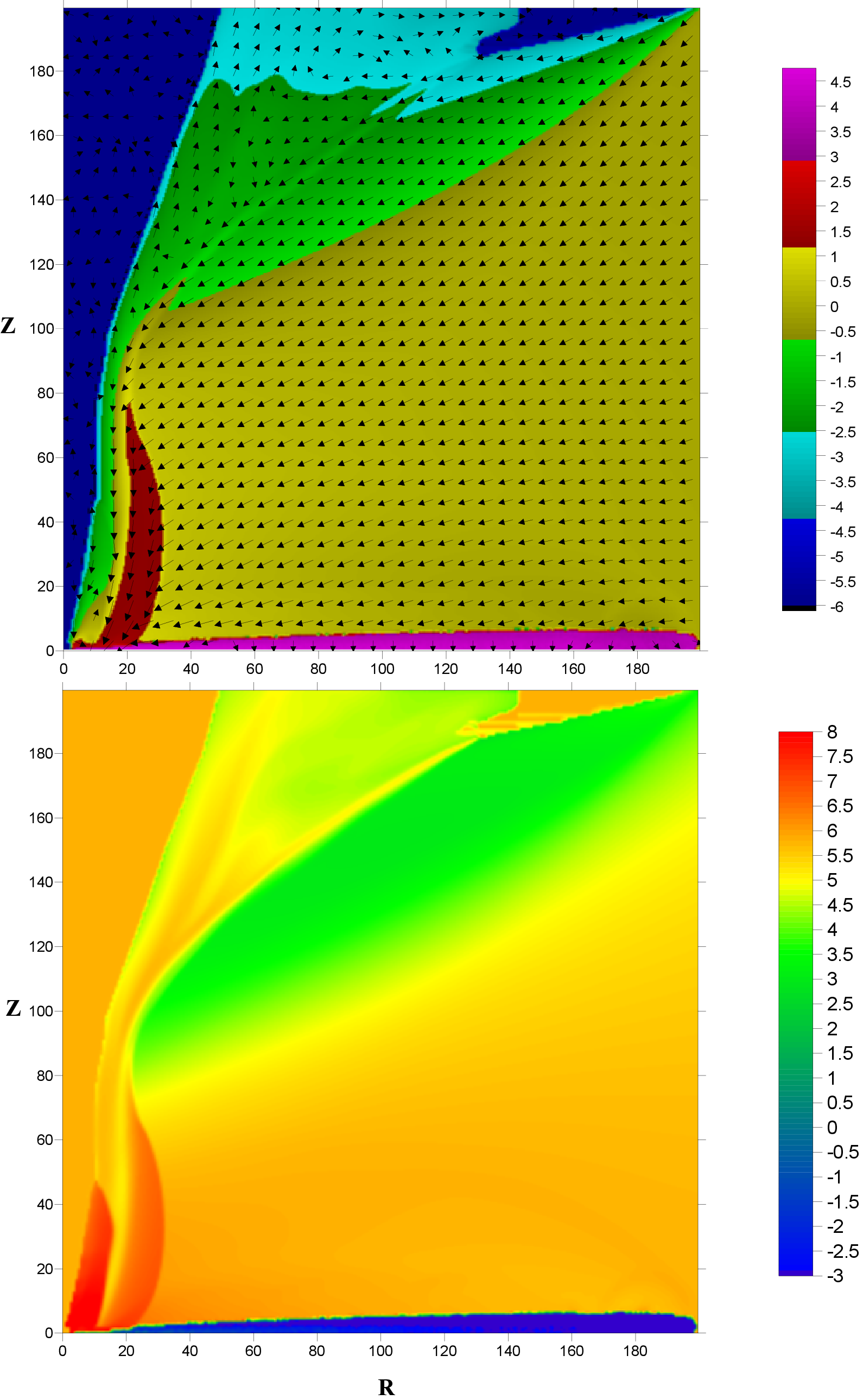}
\caption{(a) Density and velocity distributions and (b) Temperature distribution with logarithmic scale at $r-z$ plane for {\bf C1}.
Densities in normalized units are plotted in logarithmic scale as in the scale on the right.
Temperature in KeV units are plotted in logarithmic scale as in the scale on the right.
A two component flow is clearly visible.}
\end{center}
\end{figure}
Fig. 2a shows velocity and density distributions of the flow and Fig. 2b temperature distributions
in keV as per colour scale on right. Density distribution is plotted in logarithmic scale. 
The most prominent features are the formation of two component advective flow (TCAF). Both the Figures are for the case {\bf C1}.
Because of higher viscosity, the flow has a Keplerian distribution near the equatorial region.
A Keplerian disc is formed out of sub-Keplerian matter by redistributing angular momentum.
Simulation generated Keplerian disc is automatically truncated at $r \sim 15$ and a CENBOL is
produced in between the  Keplerian disk and the horizon. The region with Keplerian distribution 
becomes cooler and denser as matter settles on the equatorial plane.
Comparatively low density, sub-Keplerian and hotter matter stays away from the equatorial plane.
Liu et al. (2007) and Taam et al. (2008) investigated condensation of matter from a corona to a cool, 
optically thick inner disc.  They focused on a simple theoretical model with Compton 
cooling which leads to condensation process and maintains 
a cool inner disc. In their simulations, both components were Keplerian.
On the contrary, in our simulation, we inject sub-Keplerian flow with radial and vertical velocity
distributions at the time of injection at outer boundary. We allow the matter to advect towards black hole as  a transonic flow.
With addition of viscosity, the distribution becomes close to Keplerian only close to equatorial plane. But, we required to
add radiative cooling also to ensure that this Keplerian flow indeed radiates like a standard Keplerian disc. 
In our case, the two components rotate with two different rotational velocities and are 
therefore different from simulations by other groups. The boundary between these components is very thin and the
Keplerian disc is not at all disrupted by Kelvin-Helmholz or other instability probably because of very high
density in the Keplerian component.
\begin{figure}
\begin{center}
\includegraphics[height=8truecm,width=8truecm,angle=0]{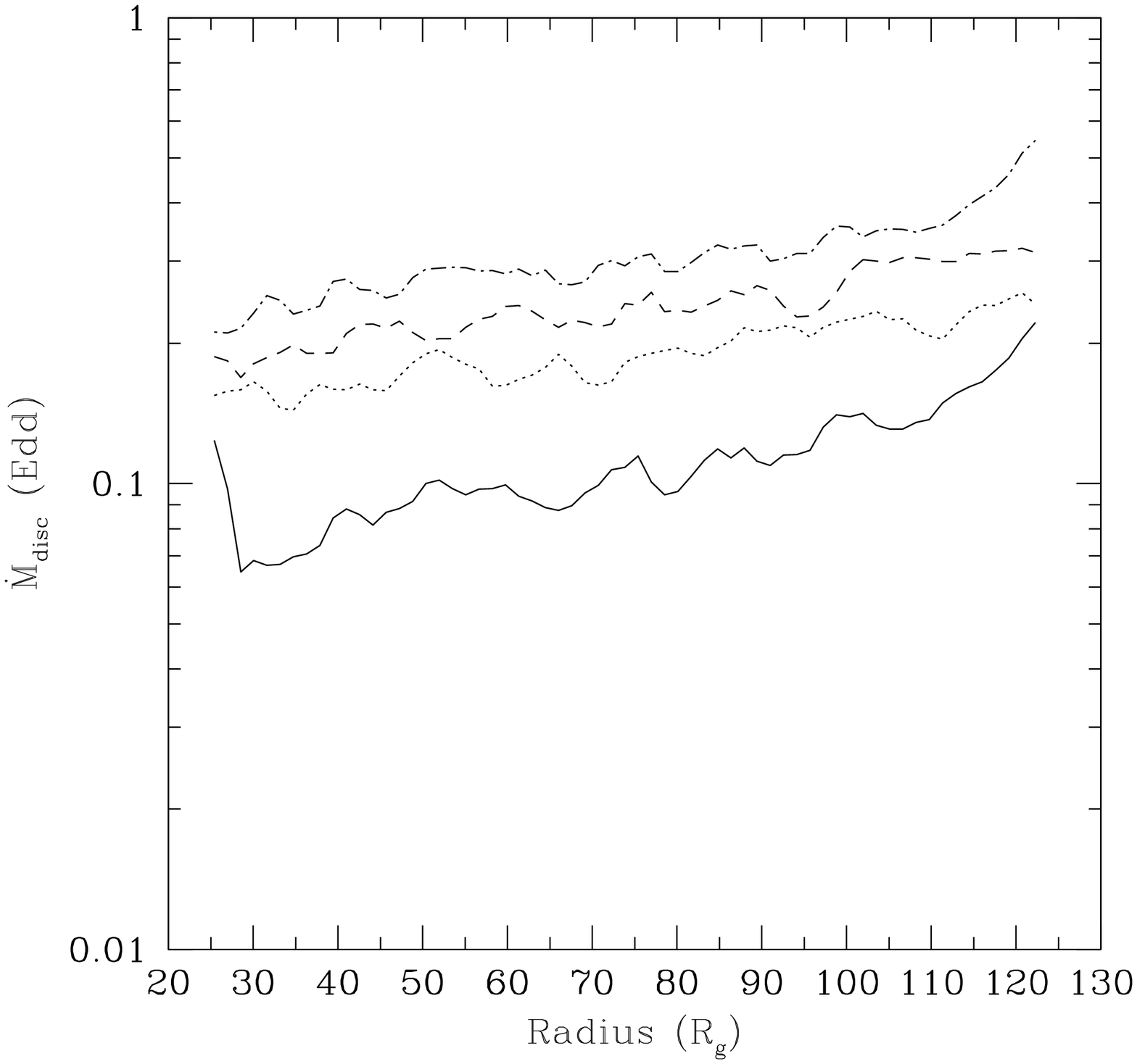}
\includegraphics[height=8truecm,width=8truecm,angle=0]{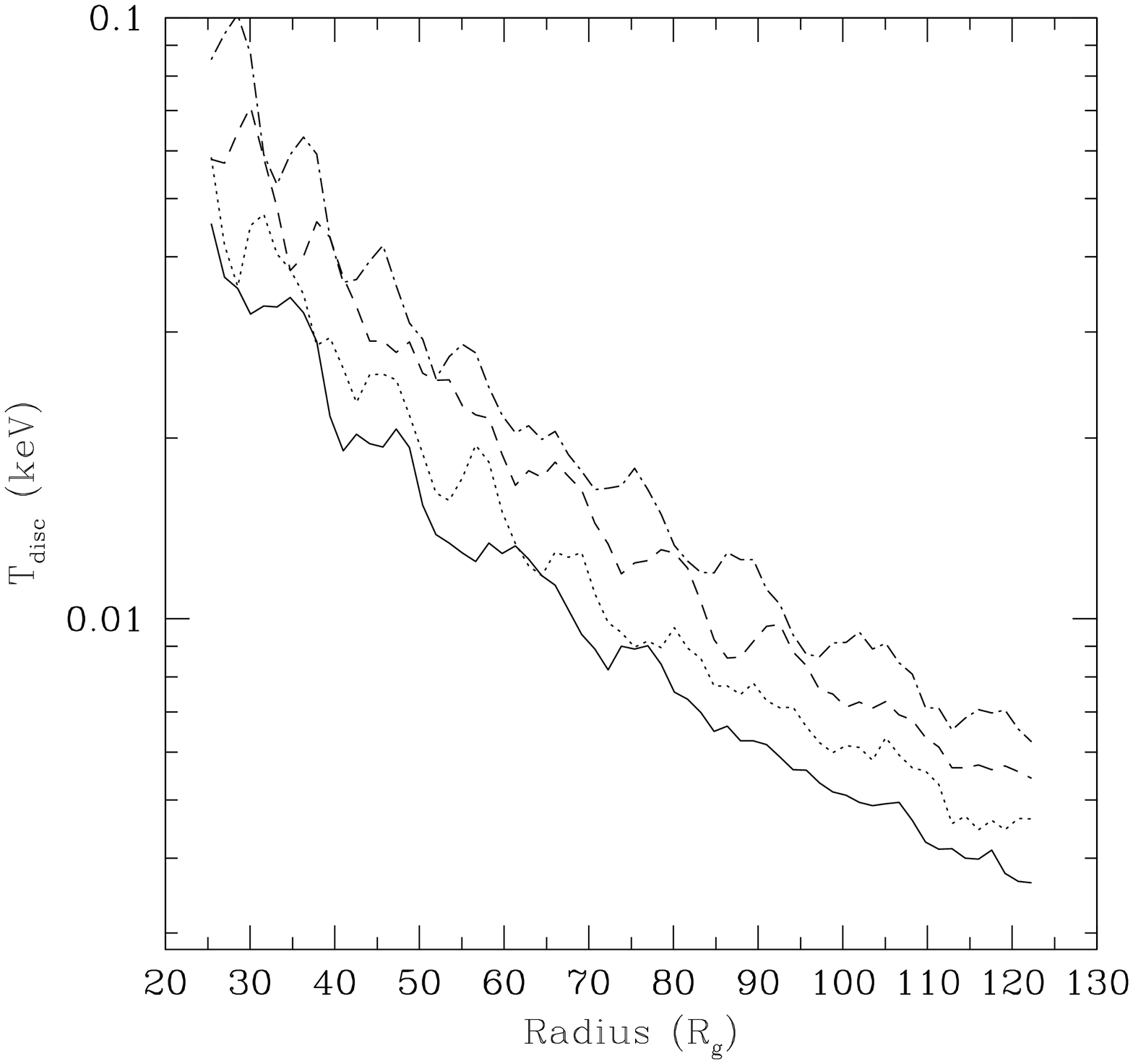}
\caption{Changes in (a) accretion rate ($\dot{M}_{disc}$) and (b) surface temperature of simulated Keplerian disc 
($T_{disc}(r)$) with change of injected accretion rate parameter $\dot{M}$ at $t= 95$ s. Here, $T_{disc}(r)$
is in keV unit, ($\dot{M}_{disc}$) is in Eddington rate unit and radius is in the Schwarzschild unit.
For both the Figures, from top to bottom, the curves are for {\bf C1, C2, C3 \& C4} respectively. 
For details, see the text.}
\end{center}
\end{figure}
In the next step, we study how the surface temperature distribution ($T_{disc}(r)$) 
and Keplerian disc accretion rate $\dot{M}_{disc}$ vary with injected accretion rate 
$\dot{M}$. One important point to note in these cases is that unlike SS73 disc 
model where $\dot{M}_{disc}$ is assumed to be constant at all radius for a 
given case, $\dot{M}_{disc}$ here appears to become radial distance dependent 
i.e., $\dot{M}_{disc} \equiv \dot{M}_{disc}(r)$. Therefore, we first calculate 
$\dot{M}_{disc}(r)$ i.e., the accretion rate of our simulated Keplerian disc in radial direction. 
At the radial distance $r$, the disc rate is obtained by $\sum (v_r \rho)2 \pi r dz$, 
where, $v_r$ is the radial velocity and $\rho$ is the mass density of the flow at $r$.
For each $r$, the $\sum$ is taken along $z$ direction over all the grid 
points from equatorial plane (i.e., $z= 0$) to disc the surface. $T_{disc}(r)$ is 
easier to compute as it is the temperature of the grid point corresponding to disc surface.
In Fig. 3, we compare radial variation of the accretion rate ($\dot{M}_{disc}(r)$) 
and surface temperatures ($T_{disc}(r)$) of our simulated Keplerian disc for
different injected accretion rates ($\dot{M}$). We have taken four cases, i.e. {\bf C1, C2, C3 \& C4} for
this purpose. To make a comparison meaningful, all runs were carried 
out up to the same time, i.e, $95$s. Values of $\dot{M}$ for which curves have been drawn are 
(from bottom to top in Fig. 3): $0.5, \  1.0, \  1.5,\  2.0$  respectively. 
It is known that local surface temperature of an SS73 disc is given by,
$$
T(r) \approx 5.48 \times 10^7 (M_{BH})^{-1/4}(\dot{M}_{disc})^{1/4} (2r)^{-3/4}
 \left[1- \sqrt{\frac{3}{r}}\right]^{1/4} \mathrm{~K},   \eqno(7) 
$$
where the disc accretion rate $\dot{M}_{disc}$ is in units of Eddington rate.
In Eqn. 7, it is evident that the surface temperature increases with Keplerian accretion rate $\dot{M_{disc}}$.
For different runs, we vary injected accretion rates $\dot{M}$ and we see that the
Keplerian accretion rate $\dot{M}_{disc}$ in simulated Keplerian disc will also vary in the same way. 
We see evidence of this in Fig. 3a. Surface temperature varies with our injected $\dot{M}$ and 
with $\dot{M}_{disc}$ which also adjusts itself with the variation of $\dot{M}$.
Surface temperature which is obtained from our simulated Keplerian disc behaves like a 
standard SS73 disc, with some quasi-periodic ripples on them. Note that the accretion rate
of the SS73-like component is not strictly constant, contrary to what is presently believed. 
:This has important implications in disc model fitting of observed data as was already pointed 
out in Dutta \& Chakrabarti (2008). The ripples are important 
and depend on viscosity parameters. This will be discussed elsewhere.
\begin{figure}
\begin{center}
\includegraphics[height=8truecm,width=8truecm,angle=0]{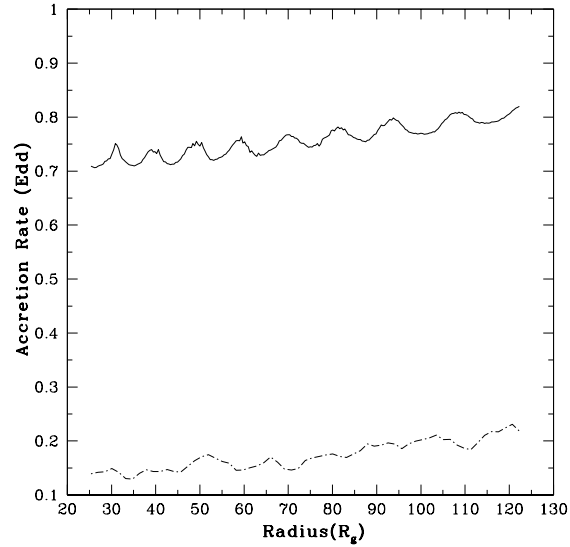}
\caption{Comparison of the disc rate $\dot{M}_{disc}$ and halo rate $\dot{M}_{halo}$ along radial direction
 in our simulation for {\bf C3} case. Here, accretion rate is in units of Eddington rate
and the distance is in units of Schwarzschild radius.}
\end{center}
\end{figure}
We have shown disc accretion rate, i.e., $\dot{M}_{disc}$, for various model runs in Fig. 3. We do the same type 
of computation for simulated sub-Keplerian hot flow. This is the so-called `halo' rate in CT95 ($\dot{M}_{halo}$).
The formula for calculating radial variation of halo rate is the same as in disc rate, 
but, the $\sum$ is taken along $z$ direction from the disc surface to the top grid. 
It is to be noted that in our simulation, there is no distinguishing boundary 
between accreting and outflowing matter. Also, some part of matter which
start as outflow from the region close to the black hole, fall back and become part of accreting halo. Therefore, it 
is difficult to compute halo accretion rate exactly. However, to draw a meaningful comparison, we compute
$\dot{M}_{halo}$ for the case {\bf C3} and in Fig. 4, we compare  $\dot{M}_{disc}$ and $\dot{M}_{halo}$ for
the same case. It is evident that average disc rate varies around $0.15$ while that of halo rate is $0.7$. 
It is understandable that the rest of the injected matter moves away at $\dot{M}_{out}$ from the system as an outflow.
\begin{figure}
\begin{center}
\includegraphics[height=8truecm,width=8truecm,angle=0]{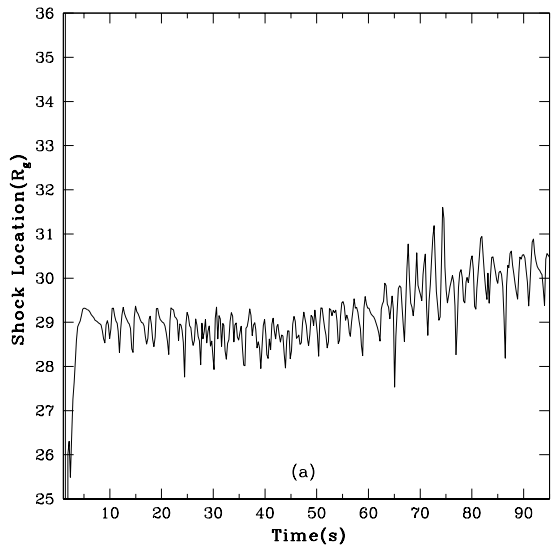}
\includegraphics[height=8truecm,width=8truecm,angle=0]{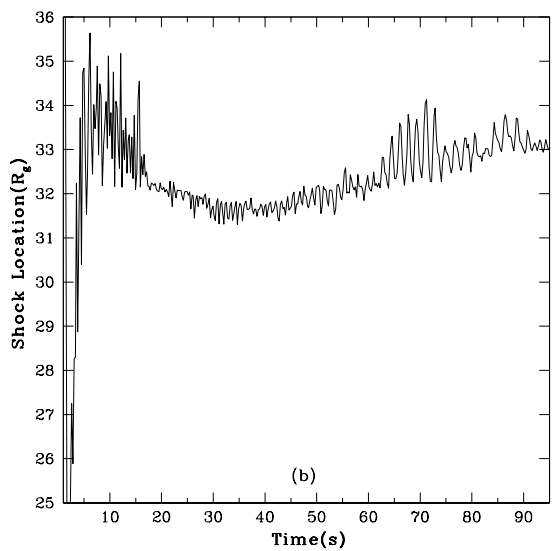}
\caption{Variation of shock location with time (in seconds)
when angular momentum is increased: (a) $\lambda=1.7$ ({\bf C1}), (b) $1.95$ ({\bf C5}).}
\end{center}
\end{figure}
\begin{figure}
\begin{center}
\includegraphics[height=8truecm,width=8truecm,angle=0]{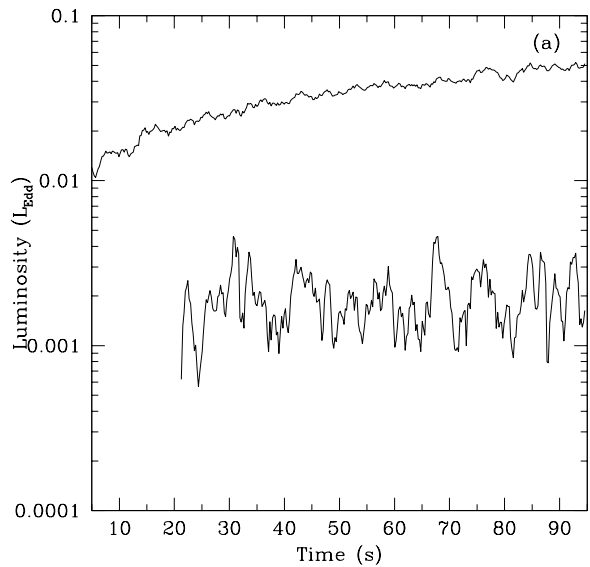}
\includegraphics[height=8truecm,width=8truecm,angle=0]{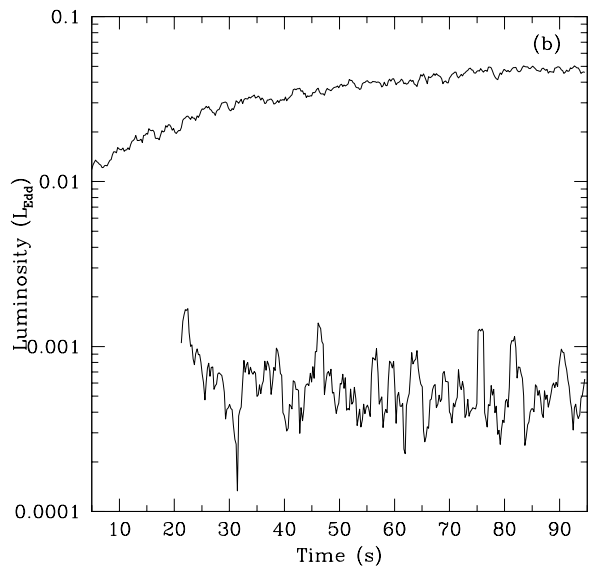}
\caption{Temporal variations of luminosity for both hot sub-Keplerian flow and Keplerian disc are shown 
when angular momentum is increased: (a) $\lambda=1.7$ ({\bf C1}), 
(b) $1.95$ ({\bf C5}). In each of the Figure, the top curve represents the hot flow while
the bottom represents the Keplerian disc. Luminosity is in units of Eddington luminosity
and time is in seconds. See text for details.}
\end{center}
\end{figure}
We now focus our attention to oscillation properties of shock locations by showing its dependence on
flow parameters. This was pointed out to be the cause of quasi-periodic oscillations or QPOs
(see, Molteni, Sponholz \& Chakrabarti, 1996; Chakrabarti, Acharyya \& Molteni, 
2004 and Garain et al., 2014 for non-viscous flows). We choose $\lambda$ = $1.7$ 
and $1.95$  at the injected flow while keeping specific energy the same. In Figs. 5(a-b), 
we compare time variation of oscillations of shock locations. Mean shock 
location increases when $\lambda$ is increased from $1.7$ ({\bf C1}) to 
$1.95$ ({\bf C5}) due to enhancement in centrifugal pressure (Chakrabarti, 1989). 
We clearly see the presence of oscillations in both the cases. 
It is interesting to study time variation of luminosity of our simulated disc.
We calculate luminosity of hot sub-Keplerian and as well as Keplerian components. There are
Total luminosity of our simulated TCAF is the sum of these two luminosities.
Radiation from hot, sub-Keplerian flow is dominated by bremsstrahlung in radiating layer 
throughout hot flow (including CENBOL) while disc luminosity is determined under the assumption of 
emission by a local black body and calculated as an integral over the Keplerian disc. 
In Figs. 6(a-b), we compare temporal variation of luminosities 
for both hot sub-Keplerian and Keplerian Components. 
Parameters are taken to be the same as before. In both the Figures, the top curve 
represents time variation of luminosity for power-law component (hot flow)
while the bottom curve shows the same for black body component, i.e., the Keplerian disc. 
Because of very large volume of the sub-Keplerian component, its luminosity 
is higher than that of the SS73 disk by a large factor. This difference would be reduced
with the increase in viscosity when more matter settles to the Keplerian component.

\section{Discussions}
The subject of accretion flow onto black holes has come a long way since the standard disc was proposed forty years 
ago. However, with the advent of precision observations, it became necessary 
to revise simplified solution obtained from a set of
algebraic equations after addition of radial velocity, viscosity and radiative transfer 
and solving global differential equations. Almost twenty years ago, 
based on such a `generalized' Bondi flow solution, (Chakrabarti, 1990, 1996),
CT95 envisaged that spectral and temporal properties of galactic and extra-galactic 
black holes may be understood if we abandon the simplest assumption that all 
accretion flows must be necessarily Keplerian. This comes about 
due to the fact that the black hole accretion disc is always transonic 
(Chakrabarti, 1990) and thus it has to deviate from a (subsonic) Keplerian 
disc anyway. Furthermore, there could be return outflow which originates 
from (low-angular momentum) inner disc and winds of companions which 
can be accreted also. In the literature, simulations of viscosity parameter 
arising out of magnetorotational instability (MRI) did not yield a 
large value of $\alpha$ parameter beyond $0.01-0.1$. This is not enough 
to transport high value of angular momentum efficiently (Arlt \& Rudiger, 
2001; Masada \& Sano, 2009). However these values are enough to transport 
low angular momentum matter and even to convert them to Keplerian discs. 
So we believe that CT95 scenario is reasonable. The belief is strengthened 
by the fact that most stellar mass black holes prefer to be in hard states 
where the SS73 disk plays a minor role.

Based on solutions of viscous transonic flows it was postulated that a transonic flow must split 
into a standard disc in regions of high viscosity parameter (near the equatorial plane), 
while regions lower $\alpha$  and inefficiently cooling matter would remain advective 
above and below the Keplerian disc. Though this configuration is widely used to fit satellite data,
so far, it illuded confirmation through rigorous numerical simulations. 
In the present paper, we show, for the first time that the standard 
disc emitting black body on the equatorial plane can indeed 
survive even when it is sandwiched between fast moving (but 
slowly rotating) sub-Keplerian flows above and below. Not only 
that, the latter component, due to inefficient loss of angular momentum, ends up producing a 
centrifugal pressure dominated shock near the inner edge, evaporating Keplerian disc (making it truncated,
so to speak) which behaves as the so-called Compton cloud, responsible 
for producing the power-law component of emitted spectrum
by inverse Comptonization of intercepted soft photons emitted from the Keplerian disc.
CENBOL is also believed to be responsible for producing jets and outflows. Its oscillation is responsible for
low frequency QPOs. We believe that with this work, stability of TCAF solution is established and 
it can be safely used for fitting spectral and temporal properties.

While we assumed that the dynamical Keplerian disc surface emits black body 
radiation, we did not include effects of radiation pressure inside the disc 
itself and used only gas pressure. This is because accretion rate inside Keplerian 
component is highly sub-Eddington (less than $0.2 {\dot m_{Ed}}$). In any case, since efficiency 
of emission for flows around a Schwarzschild black hole is only $6$ \%,  
one requires tens of Eddington rates for radiation pressure to be dynamically 
important (e.g., slim disc model of Abramowicz et al., 1988, used fifty Eddington rate as illustration). This aspect 
will be investigated in future.

Since in our work, we did not explicitly use the cooling of hots flows by the
Compton scattering, but used a power-law cooling as its proxy, it is interesting 
to compare these two cooling processes. This can be done using theoretical estimates.

Compton cooling rate for a thermal distribution of non-relativistic electrons
(hot, sub-Keplerian flow in our case) of number density $n_e$ and temperature
$T_e$ is (Rybicki \& Lightman 1979)
$$C_c = \frac{4kT_e}{m_e c^2}cn_e\sigma_T U_{rad} \mathrm{~erg~cm^{-3}~s^{-1}},$$
where, $U_{rad}$ is the radiation energy density, $\sigma_T$ is the Thomson scattering
cross-section, $c$ is the speed of light, $m_e$ is the mass of the electron and $k$
is the Boltzmann constant. We compute $U_{rad}$ from our simulation result, which is
the black body radiation that is emitted from the disk surface and enters the
sub-Keplerian flow. We can compute the energy loss ($C_b$) due to our assumed
power-law cooling term (Eqn. 1). We take the ratio at different radial distances from
the central black hole. We find that $C_b > C_c$ for $r \geq 100$;
$C_b \sim C_c$ for $r \sim r_{shock} \leq 100$; and for $r < r_{shock}$,
find $C_b << C_c$. In other words, in the post-shock region Compton cooling is much more effective
than the power-law cooling. Therefore, if we assume Compton cooling, we would expect
lesser amount of hot flow in post-shock region compared to the amount obtained
in our current simulation. Thus, hot flow rate may have been overestimated.
Absence of the explicit Compton cooling is not going affect our
conclusion of this paper, which is to say that in presence of viscosity and cooling,
stable two-component advective flow is always possible.

It is to be noted that we have not explicitly included magnetic fields, 
though, we implicitly used viscosity parameters in the general range
arising out of MRI. It is very difficult to ascertain what definite 
role magnetic fields play. However, it is clear that the presence of 
a dynamically strong poloidal magnetic field which threads both disc 
components would ensure efficient magneto-centrifugal transport of angular momentum
(Blandford \& Payne, 1982), enough to force them to act as a single component. 
However, if poloidal field is not dynamically important, two components 
will survive. What may happen is that the differential rotation would cause rapid amplification 
of toroidal component in dynamical time scale which may float out of discs 
due to buoyancy effects and collimate jets which originate from CENBOL surface
(Chakrabarti \& D' Silva, 1994).
This is outside the scope of the present paper, and will be investigated in future. 

\section{Acknowledgments}
KG acknowledges support from National Science Council of the ROC through a post doctoral
fellowship through grants NSC 101-2923-M-007 -001 -MY3 and 102-2112-M-007 -023 -MY3.
\section{Bibliography}
1.  Abramowicz, M. A., Czerny, B., Lasota, J. P. \& Szuszkiewicz, E., 1988, ApJ, 332, 646 \\
2.  Arlt, R. \& Rüdiger, G., 2001, A \& A, 374, 1035 \\
3.  Blandford, R.D. \& Payne, D., 1982, MNRAS, 199, 883\\
4.  Cambier, Hal J. \& Smith, David M., 2013, ApJ, 767, 46\\
5.  Chakrabarti, S.K. 1989, ApJ, 347, 365 (C89)\\
6.  Chakrabarti S. K., 1990, MNRAS, 243, 610\\
7.  Chakrabarti, S.K. \& D'Silva, S. 1994, ApJ, 424, 138 \\
8.  Chakrabarti, S.K. \& Titarchuk, L.G., 1995, ApJ, 455, 623 (CT95)\\
9.  Chakrabarti, S. K., 1996, Phys. Rep., 266, 229 (C96)\\
10. Chakrabarti S. K., Acharyya K. A.\& Molteni D., 2004, A \& A, 421, 1\\
11. Dutta, B. G. \& Chakrabarti, S. K., 2010, 404, 2136 \\
12. Garain, S. K., Ghosh, H., \& Chakrabarti, S. K., 2014, MNRAS, 437, 1329 \\
13. Giri, K., Chakrabarti, S. K., Samanta, M., \& Ryu, D. 2010, MNRAS 403,516\\
14. Giri, K. \& Chakrabarti, S. K., 2012, MNRAS, 421, 666\\
15. Giri, K. \& Chakrabarti, S. K., 2013, MNRAS, 430, 2836\\
16. Harten, A., 1983, J. Comp. Phys., 49, 357\\
17. Liu, B. F., Taam, R. E., Meyer, F., \& Meyer-Hofmeister, E., 2007, ApJ, 671, 695\\
18. Masada, Y. \& Sano, T. 2009,  IAU Symposium No.  259, 121\\
19. Miller, J. M., Fox, D. W., Di Matteo, T., Wijnands, R., Belloni, T., Pooley, D., Kouveliotou, C., Lewin \& W. H. G.,2001, ApJ, 546, 1055\\
20. Molteni, D., Sponholz, H. \& Chakrabarti, S. K., 1996, ApJ, 457, 805\\
21. Novikov, I. D. and Thorne, K. S., 1973, in Black Holes, ed. B. S. De Witt and
C. De Witt (New York: Gordon \& Breach), 343\\
22. Paczy\'nski, B. \& Wiita, P. J., 1980, A \& A,  88, 23\\
23. Rybicki, G. B. \& Lightman, A. P., 1979, Radiative Processes in Astrophysics (New York: Wiley-Interscience)\\
24. Ryu, D., Brown, G. L., Ostriker, J. P. \& Loeb, A., 1995, ApJ, 452, 364\\
25. Ryu, D., Chakrabarti, S. K. \& Molteni, D., 1997, ApJ, 378, 388\\
26. Shakura, N. I. and Sunyaev, R. A. 1973, A \& A, 24, 337 (SS73)\\
27. Smith, D. M., Heindl, W. A., Markwardt, C., \& Swank, J. H.,
2001b, ApJ, 554, L41\\
28. Smith, D. M., Heindl, W. A., \& Swank, J. H., 2001a, AAS,33,1473\\
29. Soria, R., Risaliti, G., Elvis, M., Fabbiano, G., Bianchi, S., \& Kuncic, Z, 2009, ApJ, 695, 1614\\
30. Soria, R., Broderick, J. W., Hao, J., Hannikainen, D.C., Mehdipour, M., Pottschmidt, K. \&
               Zhang, S.N., 2011, MNRAS, 415, 410\\
31. Sunyaev R. A., Truemper J., 1979, Nat, 279, 506\\
32. Taam, R. E., Liu, B.F., Meyer, F., \& Meyer-Hofmeister, E., 2008, ApJ, 688, 527\\
\end{document}